\documentclass[pdflatex,sn-mathphys-num]{sn-jnl}
\usepackage[a4paper, top=1in, bottom=1in, left=0.75in, right=0.75in]{geometry}

\usepackage{graphicx}%
\usepackage{multirow}%
\usepackage{amsmath,amssymb,amsfonts}%
\usepackage{amsthm}%
\usepackage{mathrsfs}%
\usepackage[title]{appendix}%
\usepackage{xcolor}%
\usepackage{tikz}
\usetikzlibrary{arrows.meta,positioning}

\tikzset{
  step/.style={
    rounded corners=2mm,
    draw=black!60,
    fill=black!5,
    inner sep=4pt,
    minimum height=7mm,
    font=\small
  },
  flow/.style={-Stealth, very thick}
}

\usepackage{textcomp}%
\usepackage{manyfoot}%
\usepackage{booktabs}%
\usepackage{algorithm}%
\usepackage{algorithmicx}%
\usepackage{algpseudocode}%
\usepackage{listings}%
\usepackage{subcaption}
\usepackage{booktabs,tabularx}
\usepackage{comment}

\theoremstyle{thmstyleone}%
\theoremstyle{thmstyletwo}%

\theoremstyle{thmstylethree}%
\newcommand{\GAM}{GAMBiT}

\newcommand{\CV}{CogVuln}
\newcommand{\MA}{MITRE ATT\&CK}

\newcommand{\CVs}{\CV{}s}
\newcommand{\Cyph}{CyPhiD}
\newcommand{\Aph}{AphiD}

\newcommand{\mt}[1]{\texttt{#1}}

\begin{document}

\title[GAMBiT Analysis]{Guarding Against Malicious Biased Threats (GAMBiT): Experimental Design of Cognitive Sensors and Triggers with Behavioral Impact Analysis}


\author[1]{\fnm{Brandon} \sur{Beltz}}\email{bbeltz@bullsrungroup.com}

\author[6]{\fnm{Po-Yu} \sur{Chen}}\email{frankchen4637@vt.edu}

\author[1]{\fnm{James} \sur{Doty}}\email{jdoty@bullsrungroup.com}

\author[2]{\fnm{Yvonne} \sur{Fonken}}\email{yvonne.m.fonken@rtx.com}

\author[3]{\fnm{Nikolos} \sur{Gurney}}\email{gurney@ict.usc.edu}

\author[6]{\fnm{Hsiang-Wen} \sur{Hsing}}\email{ghsing@vt.edu}

\author[5]{\fnm{Sofia} \sur{Hirschmann}}\email{hirschmann.so@northeastern.edu}

\author[2]{\fnm{Brett} \sur{Israelsen}}\email{brett.israelsen@rtx.com}

\author[6]{\fnm{Nathan} \sur{Lau}}\email{nkclau@vt.edu}

\author[1]{\fnm{Mengyun} \sur{Li}}\email{Mli@bullsrungroup.com}

\author[5]{\fnm{Stacy} \sur{Marsella}}\email{s.marsella@northeastern.edu}

\author[1]{\fnm{Michael} \sur{Murray}}\email{mmurray@bullsrungroup.com}

\author[6]{\fnm{Jinwoo} \sur{Oh}}\email{jinwoo@vt.edu}

\author[7]{\fnm{Amy} \sur{Sliva}}\email{amysliva@kings.edu}

\author[2]{\fnm{Kunal} \sur{Srivastava}}\email{kunal.srivastava@rtx.com}

\author[1]{\fnm{Stoney} \sur{Trent}}\email{stoney@bullsrungroup.com}

\author[2]{\fnm{Peggy} \sur{Wu}}\email{peggy.wu@rtx.com}

\author[4]{\fnm{Ya-Ting} \sur{Yang}}\email{yy4348@nyu.edu}

\author[4]{\fnm{Quanyan} \sur{Zhu}}\email{qz494@nyu.edu}

\affil[1]{\orgname{The Bulls Run Group}, \orgaddress{\street{9207 Bulls Run Pkwy}, \city{Bethesda}, \postcode{20817} \state{MD}, \country{USA}}}

\affil[2]{\orgname{Raytheon Technologies}, \orgaddress{\street{1000 Wilson Blvd.}, \city{Arlington}, \postcode{22209}, \state{VA}, \country{USA}}}

\affil[3]{\orgdiv{Institute for Creative Technologies}, \orgname{University of Southern California}, \orgaddress{\street{Playa Vista}, \postcode{90094}, \state{CA}, \country{USA}}}

\affil[4]{\orgdiv{Department of Electrical and Computer Engineering}, \orgname{New York University}, \orgaddress{\city{New York}, \postcode{10012}, \state{NY}, \country{USA}}}

\affil[5]{\orgdiv{Khoury College of Computer Sciences}, \orgname{Northeastern University}, \orgaddress{\city{Boston}, \postcode{02115}, \state{MA}, \country{USA}}}

\affil[6]{\orgdiv{Grado Department of Industrial and Systems Engineering}, \orgname{Virginia Tech}, \orgaddress{ \city{Blacksburg}, \postcode{24061}, \state{VA}, \country{USA}}}

\affil[7]{\orgdiv{Department of Computer Science}, \orgname{King’s College}, \orgaddress{\city{Wilkes-Barre}, \postcode{18711}, \state{PA}, \country{USA}}}


\abstract{This paper introduces GAMBiT (Guarding Against Malicious Biased Threats), a cognitive-informed cyber defense framework that leverages deviations from human rationality as a new defensive surface. Conventional cyber defenses assume rational, utility-maximizing attackers, yet real-world adversaries exhibit cognitive constraints and biases that shape their interactions with complex digital systems. GAMBiT embeds insights from cognitive science into cyber environments through cognitive triggers, which activate biases such as loss aversion, base-rate neglect, and sunk-cost fallacy, and through newly developed cognitive sensors that infer attackers' cognitive states from behavioral and network data. Three rounds of human-subject experiments (total n=61) in a simulated small business network demonstrate that these manipulations significantly disrupt attacker performance, reducing mission progress, diverting actions off the true attack path, and increasing detectability. These results demonstrate that cognitive biases can be systematically triggered to degrade the attacker's efficiency and enhance the defender's advantage. GAMBiT establishes a new paradigm in which the attacker’s mind becomes part of the battlefield and cognitive manipulation becomes a proactive vector for cyber defense.\footnote{Corresponding author: Ya-Ting Yang (e-mail: yy4348@nyu.edu).}\footnote{This work has been submitted to Springer for possible publication.}}




\maketitle

\section{Introduction}\label{sec:intro}

Conventional cyber defense systems have long been grounded in an assumption of rational adversarial behavior. Attackers are typically modeled as deterministic or strategic optimizers, maximizing their objectives under complete or bounded information. This abstraction has enabled decades of research in intrusion detection, automated mitigation, and adversarial modeling \cite{Kamhoua2026AutonomousCyberResilience,pawlick2021game,kamhoua2021game}. Yet, in real-world operations, particularly those involving human adversaries, decision-making is rarely so systematic \cite{aggarwal2024discovering}. Human attackers bring with them cognitive constraints, emotional stress, and perceptual biases that shape their interactions with complex digital systems \cite{huang2023cognitive}. These factors lead to systematic departures from rationality, known as \emph{cognitive vulnerabilities}, that can, paradoxically, be transformed into exploitable opportunities for defenders \cite{Zhu2025CyberDeception}. Just as traditional cyber defenses exploit vulnerabilities in code or protocols, defenders can exploit vulnerabilities in human reasoning.

The GAMBiT (Guarding Against Malicious Biased Threats) framework introduces a new paradigm for cyber defense that embeds insights from cognitive science directly into the design of cyber environments. Rather than focusing solely on the technical mechanisms of attack, GAMBiT focuses on the cognitive mechanisms that drive attacker behavior. Its central hypothesis is that human cognitive biases can be measured, modeled, and systematically influenced through carefully engineered manipulations in the environment. By embedding psychologically meaningful cues, referred to as \textit{cognitive triggers}, into networked systems, defenders can subtly induce predictable deviations from rational and commonly effectively decision-making. These deviations manifest as wasted time, misdirected effort, or noisy behavior, all of which enhance the defender’s ability to detect and counter the adversary.

The motivation for this work arises from both scientific and operational considerations. From a scientific perspective, GAMBiT aims to advance the understanding of human cognition within adversarial cyber contexts, treating the attacker not merely as an algorithmic process but as a human decision-maker with measurable cognitive attributes \cite{huang2024psyborg+}. From an operational perspective, the framework provides a method for proactive defense \cite{hu2023game,huang2020dynamic,yang2025prada}: by influencing attacker perception and judgment before exploitation occurs, defenders can preemptively degrade adversarial performance. This represents a shift from reaction to manipulation, turning the cognitive processes of the adversary into a new vector for control. Through this lens, the mind of the attacker becomes part of the battlefield and the environment becomes an instrument of psychological shaping.

The work presented in this paper integrates three major components: the design of cognitive triggers, the development of cognitive sensors, and the execution of controlled human-subject experiments to evaluate behavioral impacts. 
Cognitive triggers are engineered environmental cues embedded within a realistic enterprise network. Each cue was designed to activate a specific cognitive bias, such as loss aversion \cite{ert2013descriptive}, base rate neglect \cite{cox2020stuck}, confirmation bias \cite{katakwar2023attackers}, sunk cost fallacy \cite{hans2025quantifying}, or availability bias \cite{tversky1974judgment}. These triggers were implemented as plausible artifacts: decoy files, fake administrative accounts, aliased commands, or proxy redirections that manipulate the attacker's expectations while maintaining technical realism. Complementing these manipulations are newly developed cognitive sensors, analytic systems that translate behavioral and network data into probabilistic inferences about the attacker’s cognitive state. These include a large language model–based Attack Summarization Module (ASM) that processes Suricata alerts and NetFlow data \cite{hans2025security}, and a Theory-of-Mind Defender Agent built on the PsychSim framework \cite{Pynadath2005-ln} that models attacker biases as dynamic belief states. Together, these modules enable a defender to infer not only what the attacker is doing, but why they are doing it.

The experimental evaluation of GAMBiT involved three rounds of human-subject research conducted between July 2024 and March 2025, encompassing a total of sixty-one participants distributed across trigger and control conditions. Participants were tasked with infiltrating a simulated small business network and exfiltrating sensitive data under time constraints. The network environment consisted of thirty-eight virtual machines, with thirteen being targets and forming a valid attack path. In the trigger condition, psychologically informed artifacts were embedded at key points along this path, whereas in the control condition, the same network was presented without such manipulations or artifacts. Across all sessions, comprehensive data were collected from system logs, network telemetry, command histories, and participant self-reports. This multimodal dataset supported both quantitative behavioral analysis and validation of sensor-based cognitive inference models.

Results from these experiments reveal clear evidence that cognitive manipulations can measurably affect attacker behavior. Participants exposed to cognitive triggers exhibited reduced mission progress compared to participants in the control group, with a statistically significant difference in depth of network penetration (\emph{Kruskal–Wallis} $p=0.0495$). Participants exposed to triggers also directed a smaller proportion of their commands toward nodes on the true attack path, indicating successful diversion and resource wastage (\emph{two-way ANOVA} $p=0.003$). Certain triggers, such as Trigger 12.1 targeting loss aversion, produced particularly strong effects by enticing attackers to pursue fast but risky options, leading to increased time expenditure and misdirected effort. Moreover, networks containing cognitive triggers showed higher numbers of Suricata alerts (\emph{t-test} $p=0.0381$), suggesting that bias-driven deviations in attacker behavior made them more detectable. Collectively, these findings demonstrate that cognitive vulnerabilities can be systematically activated to degrade attacker efficiency, increase detection opportunities, and expand the defender’s informational advantage.

The remainder of this paper details the theoretical foundations, experimental design, and empirical outcomes of the GAMBiT project. Section~2 elaborates on the concept of cognitive vulnerabilities and their role in cyber operations, while Section~3 describes the experimental setup, including recruitment, network architecture, and data collection procedure. Section~4 introduces the cognitive sensors, including the LLM-based Attack Summarization Module and the Theory-of-Mind defender model. Section~5 presents the design and implementation of cognitive triggers, tracing their evolution across experimental phases. Section~6 outlines the development of the Expert Knowledge Model and Surveillance Sensors for bias detection. Section~7 presents the behavioral impact analysis, highlighting how cognitive manipulations altered attacker strategy, focus, and detectability. Finally, Section~8 concludes with reflections on the implications of cognitive cyber defense and outlines future directions for automated, psychologically informed defensive systems.

\section{Cognitive Vulnerabilities in Cyber Operations}

Traditional cyber defenses operate under the assumption that attackers are purely rational agents, optimizing their strategies based on logical understanding or reasoning of their goals, means, and environment. In such models, adversaries are treated as algorithmic optimizers, maximizing utility functions under available information and stable preferences. However, this abstraction diverges sharply from reality. Human attackers, whether professional penetration testers or state-sponsored operators, exhibit bounded rationality \cite{chen2019interdependent}: their choices are shaped by cognitive limitations, emotional states, situational stress, and habitual heuristics. In adversarial contexts characterized by uncertainty and time pressure, these factors systematically distort perception, reasoning, and judgment.

These \emph{cognitive vulnerabilities}, which are predictable deviations from normative rationality, represent an untapped dimension of the cyber battlespace. Just as traditional defenses exploit vulnerabilities in code or protocol design, cognitive defenses can exploit vulnerabilities in human decision-making. Biases such as \emph{loss aversion}, \emph{confirmation bias}, or \emph{base rate neglect} predispose attackers toward certain interpretations or behaviors when interacting with complex digital environments. By embedding psychologically informed manipulations in the cyber infrastructure, defenders can transform the environment itself into a form of adaptive deception \cite{shinde2024modeling,li2025texts,yang2025bi}, one that engages the attacker’s mind as the primary target. 

\subsection{From Cognitive Biases to Operational Triggers}

This idea underpins the GAMBiT (Guarding Against Malicious Biased Threats) framework, which operationalizes cognitive vulnerabilities through a system of embedded \textit{cognitive triggers}. These triggers are contextually grounded covert artifacts, such as fake files, credential stores, directory structures, or service configurations, strategically designed to activate specific cognitive biases. Their objective is not to deny access or generate alerts, but to subtly \emph{shape} the attacker’s perception of the environment, directing them towards inefficient, risky, or irrational actions. 

Unlike traditional countermeasures that rely on detection followed by reactive containment, cognitive triggers operate \emph{preemptively}. They exploit the internal mental model of the system of the attacker: what the attacker expects to find and how they interpret the signals along the way. By understanding attackers' heuristics or exploiting the corresponding biases, triggers introduce small, contextually believable inconsistencies that magnify into substantial deviations in strategy. In this way, deception becomes not only defensive but \emph{behavioral}: it perturbs the attacker’s reasoning loop itself and thus actions.

\subsection{Triggers as Behavioral Shapers and Implicit Sensors}

A trigger can be defined as any engineered environmental cue intended to elicit a specific bias in human cognition. Its design is guided by both technical plausibility and psychological specificity. For example, a decoy file named “passwords\_2023.txt” exploits the \emph{availability bias}—the human tendency to prioritize salient or familiar cues—while a set of misleading administrator accounts (“user-adm”) exploits \emph{base rate neglect} by prompting the attacker to assume privilege from convention rather than verification. The elegance of such triggers lies in their invisibility: they appear to be natural artifacts of the system rather than deliberate traps.

Each trigger serves a dual purpose. First, it functions as a \emph{behavioral shaping mechanism}, diverting adversaries away from optimal exploitation paths toward dead ends, delays, or noisy behaviors that increase their detectability. Second, it acts as an \emph{implicit sensor}, a cognitive probe that reveals how an attacker reasons under uncertainty. By analyzing the sequence, timing, and persistence of interactions with a trigger, defenders can infer deeper attributes of the adversary’s cognitive state: overconfidence, impulsivity, perseverance, or susceptibility to salient cues. These inferences populate a higher-level analytic system, such as the \emph{Attacker Reasoning Module (ARM)}, which estimates the likelihood that a given behavioral pattern reflects a particular bias.

Through this dual mechanism, cognitive triggers blur the distinction between \emph{defense} and \emph{intelligence collection}. They do not merely obstruct; they observe. Each interaction produces a behavioral signature that enriches the defender’s situation awareness and feeds into adaptive learning modules capable of updating the defender’s own strategy in real time.

\subsection{Cognitive Foundations and Behavioral Mechanisms}

The theoretical basis of this approach draws from cognitive psychology \cite{anderson2005cognitive}, behavioral economics \cite{kahneman2003maps}, and decision theory. Under uncertain and complex situations, humans rely on heuristic shortcuts to reduce cognitive load for satisficing solutions: rules of thumb such as ``trust administrative labels'' or ``follow the most obvious filename''. These heuristics, while efficient, introduce systematic distortions. For instance:
\begin{itemize}
    \item \textbf{Loss Aversion:} Individuals prefer to avoid losses to acquire equivalent gains; thus, attackers may take reckless shortcuts to avoid perceived setbacks in progress.
    \item \textbf{Base Rate Neglect:} Attackers overweight specific cues while ignoring prior probabilities, leading them to act on surface plausibility.
    \item \textbf{Confirmation Bias:} Once committed to a hypothesis, attackers only seek evidence to support it, even when contradictory data are available.
    \item \textbf{Sunk Cost Fallacy:} Continued investment in a failing path occurs because the prior effort psychologically demands justification.
    \item \textbf{Availability Bias:} Decision salience is driven by familiarity or ease of recall, causing attackers to focus on visible or conventional cues.
\end{itemize}

When these biases are deliberately elicited through environmental design, the defender gains an asymmetric advantage. Unlike overt deception (e.g., honeypots) that can be detected and avoided once revealed, cognitive triggers operate below conscious awareness. They influence the attacker’s perception of reality rather than the reality itself, making them more resilient to discovery and circumvention.

\subsection{Integration with Cognitive Sensing and Adaptive Defense}

The behavioral traces left by the interactions with the triggers are captured and analyzed through \emph{cognitive vulnerability sensors}. These sensors,  implemented via log analysis, network monitoring, and language-model-based summarization, convert raw activity data into probabilistic assessments of attacker bias. The resulting cognitive profile feeds into a \emph{Theory-of-Mind Defender Agent}, which models the attacker as a decision-making entity influenced by varying degrees of cognitive bias. This allows the defender to simulate counterfactual scenarios (e.g., “if I deploy this trigger, the attacker will likely choose X”) and adapt deception strategies dynamically.

Ultimately, cognitive vulnerabilities form a bridge between human psychology and cyber defense engineering. By embedding insights from behavioral science into technical systems, defenders can transform cyberspace into a psychologically adversarial environment, one where attackers are not only detected but systematically drawn into self-defeating behaviors.



\section{Experimental Design and Setup}

The results presented here span three rounds of human subjects research (Experiment 1, Experiment 2 Control Group, and Experiment 2 Trigger Group), conducted from July 2024 through March 2025. The chronologically first data collection event (EXP1) (Jul–Sep 2024) and the third data collection event (EXP2 Trigger) (Feb–Mar 2025) employed a Trigger condition, while the second data collection event (EXP2 Control) (Nov–Dec 2024) served as the Control condition. 

\begin{figure}[htbp]
    \centering
    \includegraphics[width=0.95\linewidth]{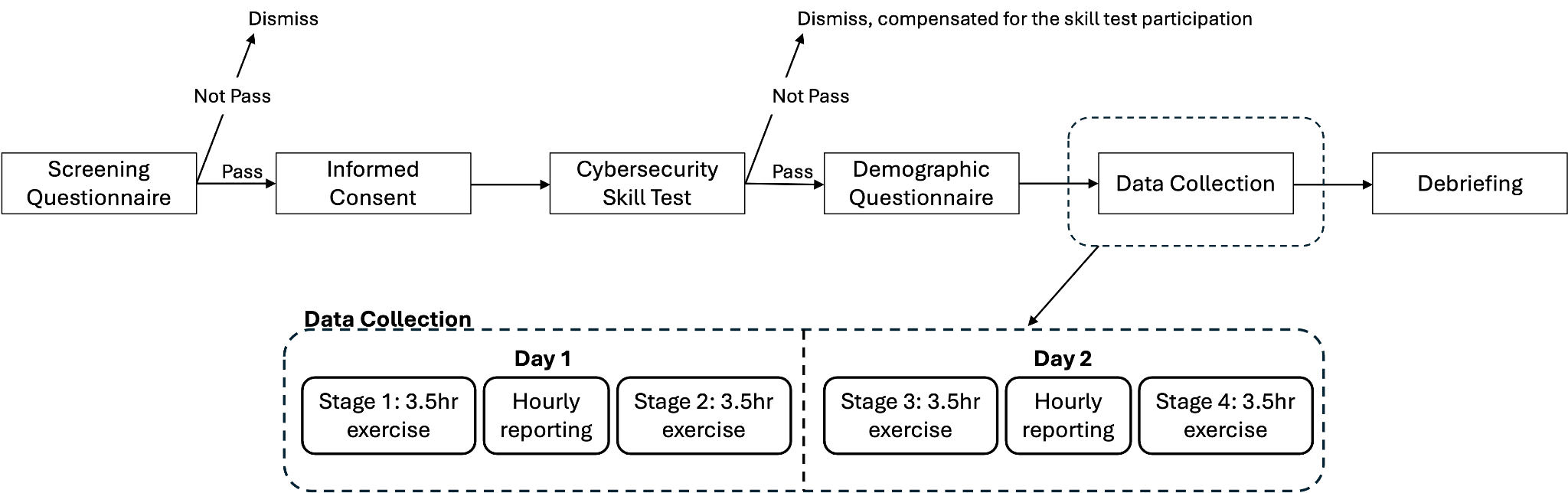}
    \caption[Flow and structure of recruitment and data collection]{Flow and structure of recruitment and data collection across the three experimental rounds (EXP1, EXP2 Control, and EXP2 Trigger). The diagram summarizes the end-to-end process from initial screening to debriefing. Participants completed a screening questionnaire, provided informed consent, and took a cybersecurity skill test that determined eligibility and division assignment. Those who passed filled out a demographic questionnaire before entering the two-day exercise. Each session comprised four hacking stages, two per day, with hourly self-reporting and continuous collection of system, network, and behavioral data. A final debriefing captured participant reflections and confirmed consistency across trigger and control conditions.}
    \label{fig:experiment_flow}
\end{figure}

\begin{figure}[htbp]
    \centering
    \includegraphics[height=7cm]{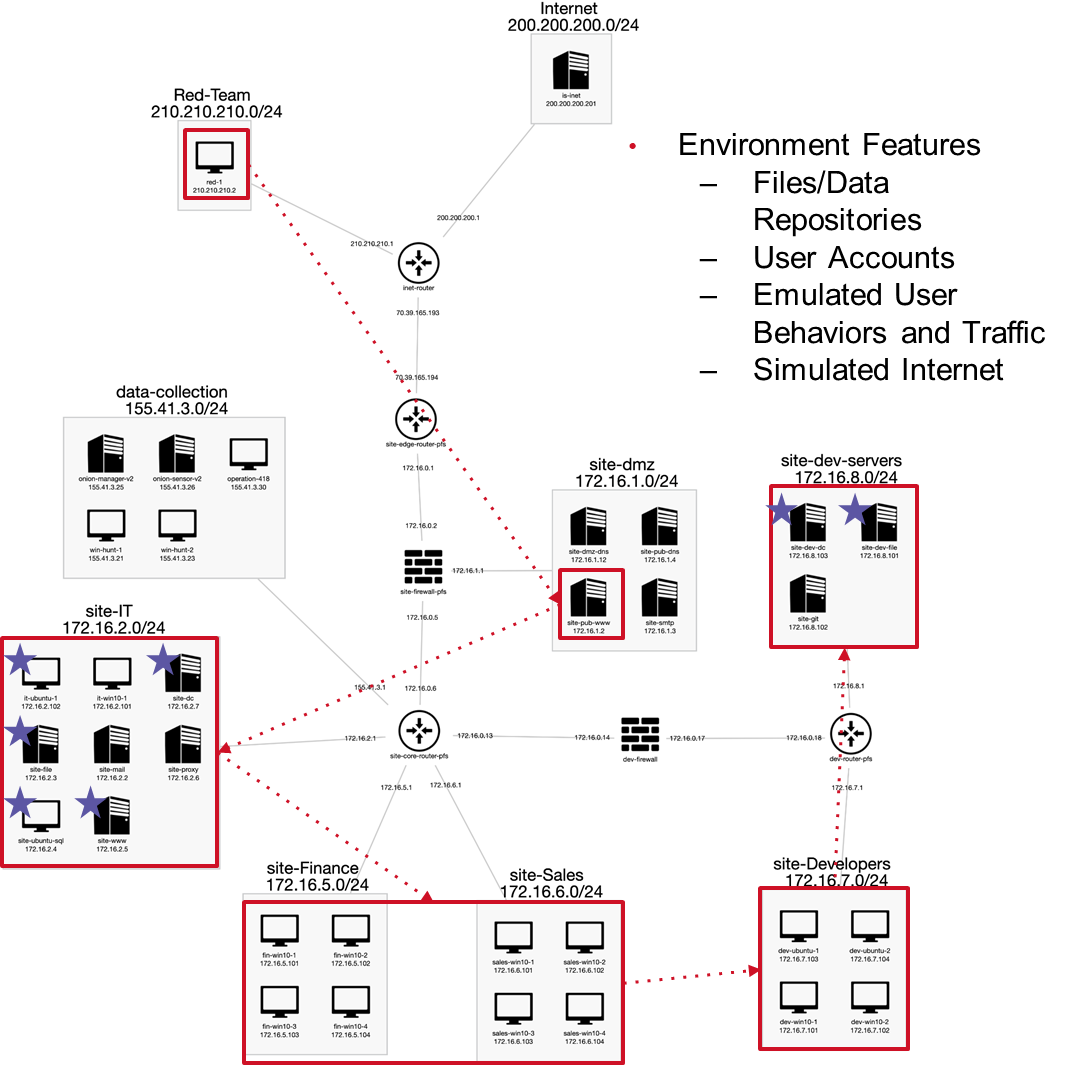}
    \caption[Network topology with shortest attack path in red]{Network topology with shortest attack path in red. Purple stars represent locations of cognitive triggers. The topology, attack path, and trigger placements are not visible to the attacker. The enterprise network emulates a small business environment distributed across multiple subnets, including IT, Finance, Sales, Developers, and Dev-Servers. Each subnet contains realistic host configurations, user accounts, data repositories, and background traffic to simulate authentic enterprise operations. The red dotted line depicts the only fully connected and exploitable sequence of hosts forming the valid attack path from the external red team network to the deepest protected targets. Cognitive triggers, such as decoy credentials, fake administrative accounts, and aliased commands, were embedded along this path to elicit measurable cognitive biases during penetration. The overall design ensured that while the system appeared operationally coherent to participants, the deceptive artifacts were indistinguishable from legitimate network elements, thereby preserving ecological validity during human-subject testing.}
    \label{fig:cyber_range}
\end{figure}

In all three experiments participants were told to infiltrate a small business network and exfill valuable data. An overview of this network is shown in Figure~\ref{fig:cyber_range}. Purple stars indicate the locations of triggers, which were absent in the control group.  The recruitment, exercise design, and data collection procedures used in EXP2 Control and EXP2 Trigger were nearly identical to those employed in EXP1. 
The two-day hacking exercise followed the same four-stage structure, with embedded triggers, event scoring, and structured check-in/check-out procedures. Data collection methods, including system logs, network traffic, and participant questionnaires, largely remained consistent across experiments, with engineering improvements made for EXP2 based on lessons learned from EXP1.  
Figure~\ref{fig:experiment_flow} outlines the overall experiment process including recruitment and the structure of the two day hacking sessions.

Table~\ref{tab:recruitment_numbers} documents the numbers throughout the recruitment process across experiment groups. As Control and EXP2 were recruited and screened together the numbers were combined for the 2 groups. In summary, 19 participants completed the two-day experiment for EXP1, 20 completed the experiment for EXP2 Control, and 22 completed the experiment for EXP2 Trigger.

\begin{table}[htbp]
\centering
\caption{Recruitment numbers across experiment groups.}
\label{tab:recruitment_numbers}
\footnotesize
\setlength{\tabcolsep}{6pt}
\begin{tabularx}{\textwidth}{l*{7}{c}}
\toprule
\multirow{2}{*}{Experiment} & \multirow{2}{*}{Interested} & \multicolumn{2}{c}{Initial screening} & \multicolumn{2}{c}{Skills challenge} & \multirow{2}{*}{Scheduled} & \multirow{2}{*}{Completed} \\
\cmidrule(lr){3-4}\cmidrule(lr){5-6}
 & & Completed & Passed & Completed & Passed & & \\
\midrule
EXP1 (Trigger)  & 96  & 83  & 56  & 45  & 27  & 25 & 19 \\
Control         & 238 & 186 & 147 & 103 & 73  & 64 & 20 \\
EXP2 (Trigger)  & 238 & 186 & 147 & 103 & 73  & 64 & 22 \\
\bottomrule
\end{tabularx}
\end{table}

\section{Cognitive Vulnerability Sensors}

Although cognitive triggers represent the active component embedded in the environment of the GAMBiT framework, their effectiveness depends on the defender’s ability to observe, measure, and interpret the resulting behavioral responses. To transform attacker interactions into actionable cognitive intelligence, the project developed a suite of \emph{Cognitive Vulnerability Sensors} (CogVuln Sensors) capable of inferring latent cognitive states from observable cyber behaviors. These sensors constitute the perceptual and analytic layer of GAMBiT’s cognitive defense architecture, providing the bridge between raw data and higher-order reasoning about adversarial cognition.

The core principle underlying these sensors is that cognition leaves a behavioral trace. Choices made under uncertainty, whether to verify a privilege, retry a failed exploit, or pursue a risky shortcut, reveal systematic patterns that can be statistically associated with underlying cognitive biases. By capturing, preprocessing, and classifying these behavioral signals, the sensors estimate the probability that an attacker is operating under specific cognitive vulnerabilities such as loss aversion, base-rate neglect, confirmation bias, sunk-cost fallacy, or availability bias. In this sense, the sensors serve a dual role: they function as measurement instruments that quantify bias tendencies, and as inference engines that update the defender’s internal model of the attacker in real time.

Section 4 details the design and implementation of this sensing architecture. It begins with a description of the CogVuln sensor pipeline, which integrates network-level telemetry and intrusion-detection alerts through a customized Large Language Model–based \textit{Attack Summarization Module} (ASM). The ASM converts heterogeneous data streams into structured \textit{MITRE ATT\&CK Technique Signals}, which are then processed to estimate attacker bias probabilities. Subsequent subsections describe the development and validation of the ASM classifier, the mathematical formulation of the CogVuln belief-updating process, and its integration into the Theory-of-Mind Defender Agent built on the PsychSim social-simulation framework. Together, these components operationalize the detection of adversarial cognition, transforming behavioral noise into structured information about how human attackers perceive, decide, and adapt within deceptive cyber environments.

\subsection{CogVuln Sensor description}
\begin{figure}[htbp]
    \centering
    \includegraphics[width=0.95\linewidth]{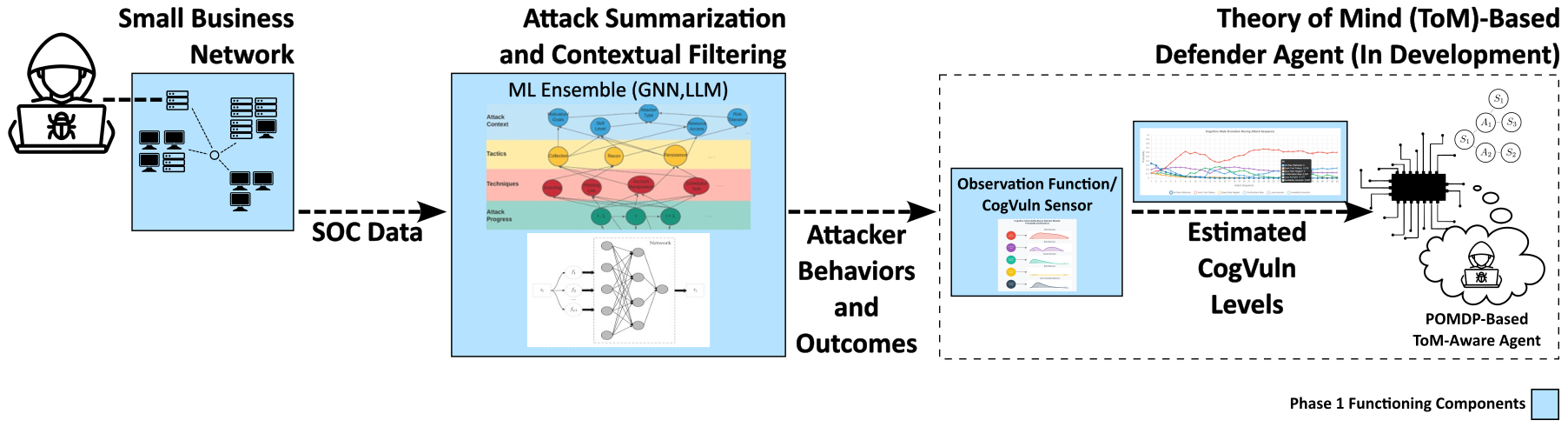}
    \caption[Pipeline describing how the ASM extracts meaningful attacker behaviors (\MA{} TTP's)]{Pipeline illustrating how the Attack Summarization Module (ASM) extracts and contextualizes attacker behaviors (MITRE ATT\&CK TTPs) to inform the GAMBiT CogVuln sensor. SOC data from the small business network, comprising Suricata alerts, NetFlow, and host logs, are processed through the ASM, which integrates an ensemble of machine learning models, including rule-based mappings and a customized large language model. These outputs are filtered with SME-informed heuristics to associate each TTP with contextual factors such as task success, target host, and timing. The resulting structured signals are then passed to the CogVuln sensor, which estimates the likelihood that an attacker exhibits specific cognitive vulnerabilities. These estimates feed into a Theory-of-Mind (ToM) defender agent that models attacker decision-making and supports adaptive deception strategies. Blue elements denote components completed; dashed boxes indicate planned extensions under development.}
    \label{fig:Sensor-Flow}
\end{figure}

Figure~\ref{fig:Sensor-Flow} shows an overview of the \CV{} sensor process. The right hand side depicts how the \CV{} sensor is integrated in a larger Theory of Mind-based Defender Agent model.  The \CV{} sensor uses preprocessed Suricata alerts and NetFlow data as input, and outputs the probability of an attacker’s vulnerability to each \CV{} (i.e., Base Rate Neglect, Loss Aversion, Availability, Confirmation, and Sunk Cost). The sensor pipeline includes a customized LLM-based Attack Summarization Module that inputs Suricata alerts and outputs timestamped \MA{} Technique Signals (MATS). Using netflow data and cybersecurity SME-informed heuristics, the MATS are linked to additional contextual information (for example, whether the task was successful or not). The combination of MATS and task contextual data is subsequently used as input to a sensor, which outputs \CV{} probabilities.

\subsection{LLM Attack Summarization Module (ASM) Description}
The LLM-ASM classifier development was a multi-step process. We first defined a subset of \MA{} TTPs relevant to our experiments, which we call \MA{} Technique Signals (MATS). This was necessary as not all TTPs were relevant, and a subset was needed to reduce the subject’s reporting burden. Within participant self-reports, subjects were asked to select the MATS that they attempted during each session of their experiment. To build the LLM ASM, we first used the self-reported MATS to identify timestamps of relevant Suricata rules that appeared in the Suricata data stream from experiments. We then consulted with SMEs to verify that the identified Suricata rules are indeed associated with the MATS. Lastly we provided this mapping to GPT-4, and subsequently to GPTmini, for processing participant data.  The LLM-ASM component generated timestamped MATS for processing by the \CV{} sensor. We audited the data to evaluate and correct for LLM hallucinations. This was done by randomly selecting data from 10 participants from experiment 1, and fine-tuning the mapping until the auditing of a third of the previously audited participants did not reveal any hallucinations.     
Note that the \GAM{} team also developed a generative neural network (GNN) ASM approach, which was not used in the pipeline to \CV{} sensor presented here. The LLM and GNN ASM approaches were planned to be combined in an ensemble approach in the future.

\subsection{Defender Agent's \CV{} Model of the Attacker}
The \GAM{} ToM Sensor leverages Psychsim \cite{Pynadath2005-ln}, a social-simulation architecture consisting of an open-source library of algorithms for modeling social cognition. It has been used for large-scale simulations of urban populations (e.g., hurricane response, patterns of life, terrorist attacks, supply chain decision making) and small-scale human-machine teaming (e.g., search-and-rescue) \cite{pynadath2023improving, doroudi2018integrated, doroudi2020effects, pynadath2023disaster, pynadath2016semi}. The \GAM{} ToM sensor is an instantiation of Psychsim to infer potential cognitive biases or tendencies in an attacker that is attacking a network, regardless of the presence or absence of triggers.
The model infers the distribution of \CVs{} of the attacker. It is an agent-based system that can have alternative, explicit mental models representing potential attacker \CV{} and an additional defender agent.  Each \CV{} model has a utility function that informs its decision-making, based on what actions would achieve its goal if under the influence of its respective \CV{}.  The defender agent observes the behavior of an attacker and formulates beliefs about the likelihood of each \CV{} model. The output of the model is a probability distribution representing the beliefs of the defense agent about the susceptibility of a participant \CV{} given the history of observations. It is possible that an agent is exhibiting multiple \CVs{} simultaneously, thus the total probability of all the \CVs{} have been normalized to sum to 1. Because these beliefs about the attacker are represented in terms of different utility functions, this will, going forward, allow a defender agent to formulate a defensive strategy. Specifically, a defender agent could simulate the attacker so as to deceive/manipulate by performing hypothetical/counterfactual reasoning of the form: ``if I take these actions, my best model of the attacker suggests they are likely to respond by taking these other actions''. 

\textbf{Data Sources}: The input to \GAM{} Psychsim is designed to take abstracted behaviors derived from human subjects, and is agnostic to the presence of triggers as well as the network topography. The inputs are the outputs of the LLM-ASM and task success data as described above.  

\textbf{Model Development Methods}: We have models of 5 biases (loss aversion, base rate neglect, confirmation bias, sunk cost fallacy, and availability bias) used to infer potential bias tendencies in the hackers. In formulating these biases, various adaptations were made based on subject matter expertise. We will provide examples of how we use the bias descriptions to make the various adaptations in the sections below. 
How specific CogVulns are instantiated in the model using TTP and contextual inputs is described per CogVuln below.

\textbf{Loss Aversion}:
General loss aversion is interpreted as the sense of preferring less risky actions \cite{o2018modeling, davies2007behavioural, ert2013descriptive}. The model approach focuses on general loss aversion. This is instantiated in the model by the assumption that the attacker prefers less risky actions related to the risk of being discovered. For belief updating, the model evaluates the risk of detections associated with each TTP relative to the maximum risk in the current state. For example, establishing persistence mechanisms would increase the risk of discovery, whereas not establishing persistence mechanisms does not increase the risk of discovery, even though it does increase the risk of not being able to reestablish access if access is lost.  These TTP risk factors come from a model of the underlying probability of detection that each action presents. The probability of detection for each TTP was initialized based on SME evaluations, but requires further refinement with additional research. TTP's with high relative risk of detection will decrease the Loss Aversion belief state, whereas TTP's with low relative risk of detection will increase the LA belief state. 

\textbf{Base Rate Neglect}:
We treat base rate neglect as the ignoring or under-weighting the prior probability (base rate) and over-weighting new evidence \cite{bar-hillel_efficient_2008}. Accordingly, the behavioral signature of base rate neglect is modeled as an attacker’s over-reaction to the recent outcome. Instead of consistently updating their prior beliefs by weighing every observation uniformly, their next move is thus more dependent on whether their prior actions succeeded or failed. The reactive condition is met if the attacker repeats a success or abandons a failure. The model checks if the attacker repeated an action (TTP) that was just successful, or if they switched to a new action (TPP) after the previous one failed. If either is true, the base rate neglect score increases. If the attacker shows persistence by repeating a failed action or explores a new one after a success, the score decreases.  The Base Rate Neglect belief update is therefore agnostic to specific TTP's, but rather evaluates task-switching in the context of successes and failures. 

\textbf{Confirmation Bias}:
We modeled the behavioral signature of Confirmation bias as an attacker’s tendency to stick with an action they believe should be successful, even when it repeatedly fails, essentially discounting that evidence  \cite{Baker2022Confirmation}. The model assesses this bias by checking if an action’s observed success rate is lower than a preset belief. Essentially, if an action is performing worse than the attacker's prior belief, and the attacker uses it anyway, the confirmation bias score increases. The score increases more when the action is performing very poorly. If the action’s success rate meets or exceeds the prior belief, the bias score is slightly lowered. For these belief updates, TTP's are evaluated against a baseline success rate determined by SMEs. 

\textbf{Sunk Cost fallacy}:
The Sunk Cost belief state is updated based on repeated target engagement and estimated future value. The implementation considers host value, difficulty, and investment already made. The hacker’s decision to continue pursuing a target is modeled by adapting an approach that is often used to model sunk cost mathematically \cite{kleinberg2021stochastic, arkes1985psychology}. Specifically, the hacker continues to pursue compromising a host if $ Expected Reward \geq -\lambda \cdot Sunk Cost$, where parameter $\lambda \geq 0$ expresses the degree of the bias.
We have adapted it here to include additional factors such as the Attacker’s estimation of host difficulty. If a repeated action is observed where the estimated future value is negative, indicating that the expected future costs are greater than or equal to any potential reward, the model updates its belief in the sunk cost bias using the calculated ratio. The estimate of sunk cost is calculated using the reward that an attacker associates with an action and the potential cost of taking said action, which is derived from the risk of discovery. The defender's belief update uses a standard reinforcement learning approach where every time the attacker takes an action that is indicative of a sunk cost vulnerability, the defender incrementally updates its beliefs about the attacker's sunk cost vulnerability by a fixed weighting factor and the degree of irrationality of the action. 

\textbf{Availability}:
The belief state for Availability is updated using external lexical analysis when target names are available, with fallback to simple increase when lexical is unavailable. The lexical analysis seeks to assess attempts on target/account names that suggest administrative privileges, with higher scores for `admin' or `root' compared to personal names. This assumes attackers’ perceptions are affected by what is familiar and expected, due to availability bias \cite{yuill2007psychological}. Of course, this captures only some facets of availability and will overestimate the likelihood of the bias given that naming of hosts and user accounts often do signify their security relevance. For example, admin accounts in reality often do have the string `admin' in their account names; indeed, this tendency may further suggest a general lack in best security practices, so it is rational to target such accounts. Nevertheless, the hacker targeting such host/accounts still suggests a defensive counter-attack approach. Our model currently works by assessing the similarity of the names of user accounts and hosts being accessed by the hacker to key words such as admin and master.

\begin{figure}[htbp]
    \centering
    \includegraphics[height=7cm]{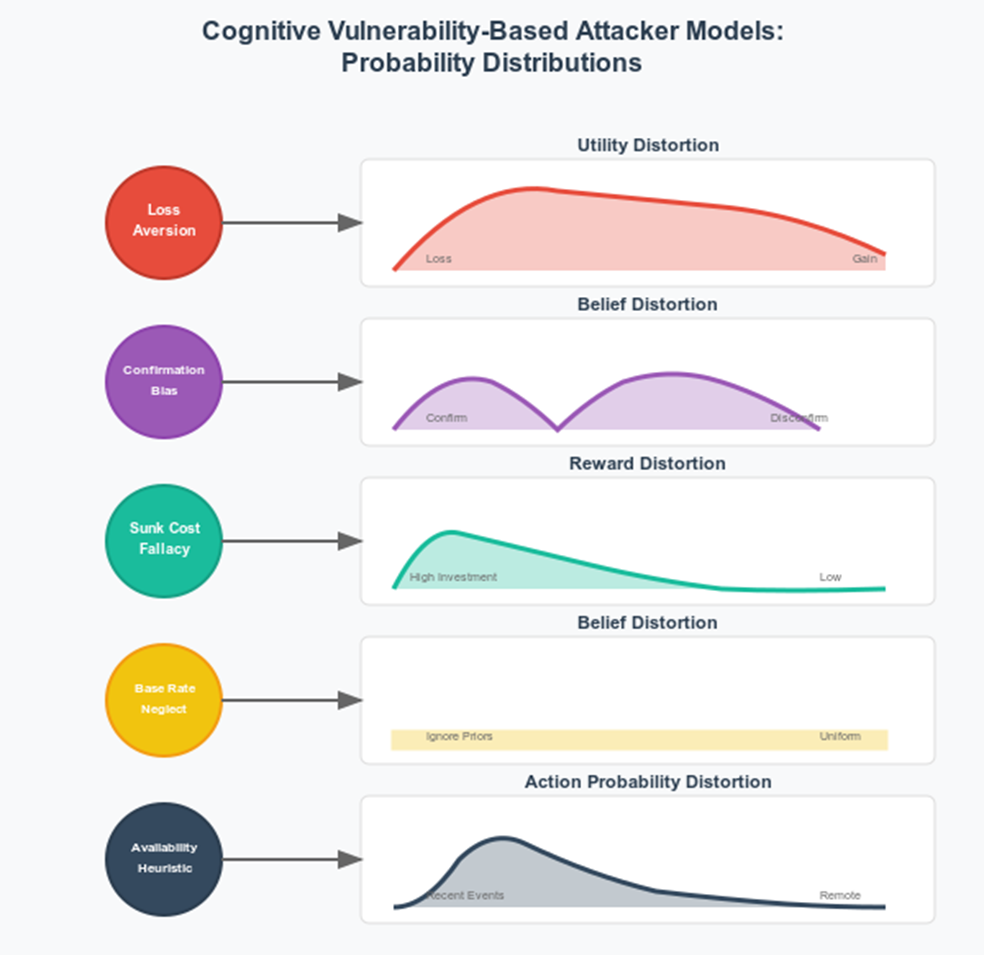}
    \caption[Illustration of probability distributions for the five \CVs{}]{Illustration of probability distributions for the five cognitive vulnerabilities (CogVulns) estimated by GAMBiT’s Theory-of-Mind (ToM) sensor. Each curve represents a distinct form of cognitive distortion that modifies how an attacker perceives value, evaluates evidence, or selects actions. Loss Aversion produces an asymmetric utility function where losses are overweighted relative to equivalent gains, biasing attackers toward low-risk, low-reward actions. Confirmation Bias manifests as a bimodal belief distribution reflecting selective weighting of evidence that supports prior assumptions while discounting contradictory signals. Sunk Cost Fallacy appears as a skewed reward distribution in which continued investment in high-cost, low-return actions is irrationally preferred due to prior effort. Base Rate Neglect flattens the belief distribution, indicating a failure to incorporate prior probabilities and an overemphasis on recent or salient outcomes. Availability Bias shapes an action probability distribution concentrated around recent or familiar cues, such as frequently observed filenames, accounts, or directories, while undervaluing remote or less salient options.}
    \label{fig:sensor_prob_dists}
\end{figure}

\textbf{Model Outputs}: Model outputs are probability distributions representing the estimated likelihood of the hacker having different biases. For visualization purposes the distributions are plotted in Figure~\ref{fig:sensor_prob_dists}. To provide a bounded input to the ToM-based Defender, the \CV{} outputs would be normalized to sum to 1, as noted earlier.


\section{Cognitive Vulnerability Triggers}

\begin{figure}[htbp]
    \centering
    \includegraphics[width=0.95\linewidth]{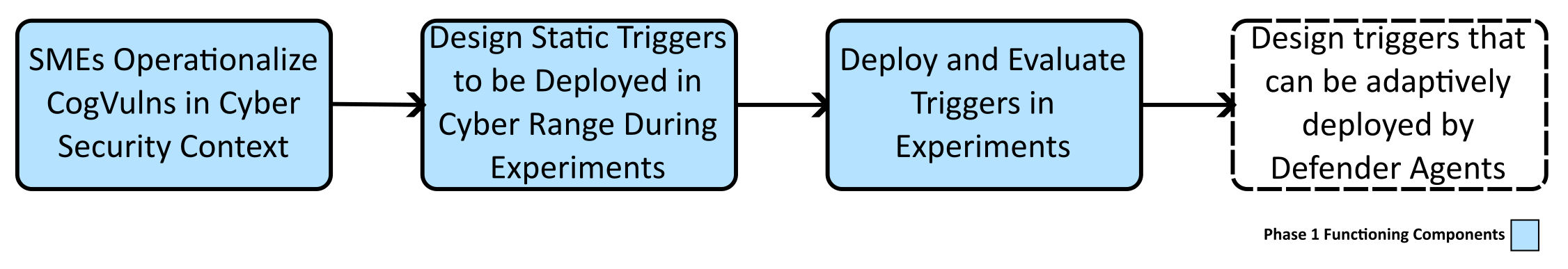}
    \caption{High-level trigger design process. The figure illustrates the end-to-end workflow for developing cognitive triggers used in the GAMBiT experiments. Subject-matter experts (SMEs) first operationalized cognitive vulnerabilities (CogVulns) within cybersecurity contexts, translating psychological constructs such as loss aversion or confirmation bias into observable decision points for attackers. Based on these mappings, the team designed static, technically plausible triggers—such as decoy credentials, fake administrative accounts, or aliased commands—to be deployed across the cyber range during human-subject experiments. Each trigger was embedded within realistic network environments to elicit specific bias-driven behaviors under naturalistic conditions. The triggers were then iteratively deployed and evaluated to measure behavioral effects and refine design parameters. While current work focused on static deployments, future work aims to develop adaptive triggers that can be dynamically selected and deployed by autonomous defender agents as part of real-time cognitive defense strategies.}
    \label{fig:trigger_process}
\end{figure}

\begin{figure}[htbp]
    \centering
    \includegraphics[width=1.00\linewidth]{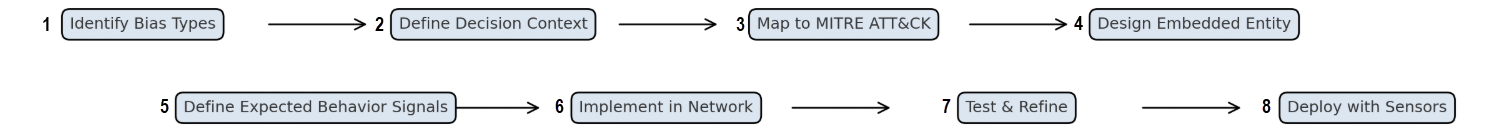}
    \caption{Detailed trigger design methodology. This figure outlines the eight-step procedure used to transform abstract cognitive biases into actionable and testable network artifacts. The process begins by identifying the target bias type and defining its associated decision context: what choice the attacker must make and under what uncertainty. Each context is then mapped to relevant MITRE ATT\&CK tactics and techniques, ensuring operational realism and analytical traceability. Next, an embedded entity is designed (e.g., a file, account, or configuration) and paired with clearly defined behavioral indicators that distinguish rational from biased responses. These triggers are implemented within the network environment, tested and refined through pilot studies, and finally deployed with integrated sensors for behavioral capture and post-experiment analysis. This structured approach ensures that every trigger is both psychologically valid and technically coherent within the broader experimental infrastructure.}
    \label{fig:trigger_detail}
\end{figure}

Figure~\ref{fig:trigger_process} depicts the design of the triggers. 
Triggers are developed to create realistic, in-range scenarios for testing how specific cognitive vulnerabilities might influence attacker decision-making. The trigger design process balanced technical plausibility with cognitive science objectives. Figure~\ref{fig:trigger_process} depicts the design of the triggers. For each trigger, the team identified the primary cognitive bias, defined the attacker-facing decision context, and mapped it to relevant \MA{} tactics and techniques. They then designed and embedded the trigger into the network as a plausible file, credential, service, or configuration, defined the expected behavioral signals, and implemented it with integrated sensors. Each trigger was tested, refined, and deployed to ensure both technical realism and its potential to elicit the intended bias. Figure~\ref{fig:trigger_detail} illustrates this process.

\subsection{Trigger evolution from Experiment 1 to Experiment 2}
Experiment 1 showed that while many triggers were valid in design, some were rarely encountered because they were placed in peripheral or low-traffic areas of the network. For Experiment 2, the team increased the likelihood of encounters by relocating several triggers to more prominent network locations along common attack paths. In addition, new data sources, such as expanded host logs and additional Suricata alerting, were integrated to improve post-experiment analysis of trigger interactions. These changes aimed to both raise encounter rates and enrich the dataset for behavioral interpretation.

Across both experiments, the core structure and intent of the triggers remained consistent, but Experiment 2 featured more diverse trigger types, refined implementations, and broader coverage of attack phases. The five most frequently encountered triggers are summarized in Table~\ref{tab:triggers}. 
Triggers were labeled using a structured code combining bias type, trigger class, and instance hierarchy. The first letter indicates the intended cognitive bias (e.g., B = Base Rate Neglect, L = Loss Aversion). The first number denotes the trigger’s class or type, as described in the Cognitive Vulnerability Playbook. The second and third numbers, if present, identify specific instances of that trigger type when multiple versions were deployed in different network locations. This system allowed rapid identification of both the bias focus and the technical role of each trigger.


\begin{table}[htbp]
\caption[Example trigger conditions and associated cognitive biases.]{Example trigger conditions and associated cognitive biases. The final four columns show the number and percentage of participants that encountered and interacted with a given trigger. This number is based on the amount of participant who progressed far enough to have interaction with the trigger. (E.=Encountered; I.=Interacted)}
\label{tab:triggers}
\centering
\begin{tabular}{p{1.1cm}|p{1.5cm}p{2.4cm}p{1.25cm}p{1.25cm}p{1.2cm}p{1.2cm}}
\hline
\textbf{Trigger ID} & \textbf{Desc.} & \textbf{Bias Type} & \textbf{\# P's E.} & \textbf{\% P's E.} & \textbf{\# P's I.} & \textbf{\% P's I.} \\
\hline
S.9.4 & Password-protected decoy & Sunk Cost  & 18 & 95\% & 10 & 56\% \\
C.7.1.1 & Aliased commands loop & Confirmation, Sunk Cost  & 14 & 74\% & 8 & 57\% \\
B.2.1.1 & Fake admin accounts & Base Rate Neglect & 13 & 68\% & 9 & 69\% \\
L.12.1 & Proxy redirect & Loss Aversion & 17 & 89\% & 12 & 71\% \\
A.3.1.1 & Salient filenames & Availability & 10 & 53\% & 6 & 60\% \\
\hline
\end{tabular}
\end{table}

\subsection{Example triggers}
For illustration purposes, two triggers are described in detail in this section. Triggers 2.1.1 and 12.1 were two of the five triggers that were most encountered, and are shown to be some of the most successful in terms of behavioral impacts (see Chapter~\ref{ch:results})

\subsubsection{Trigger 2.1.1 (Base Rate Neglect) }
\begin{figure*}[htbp]
    \centering
    \begin{subfigure}[t]{0.48\textwidth}
        \centering
        \includegraphics[height=3cm]{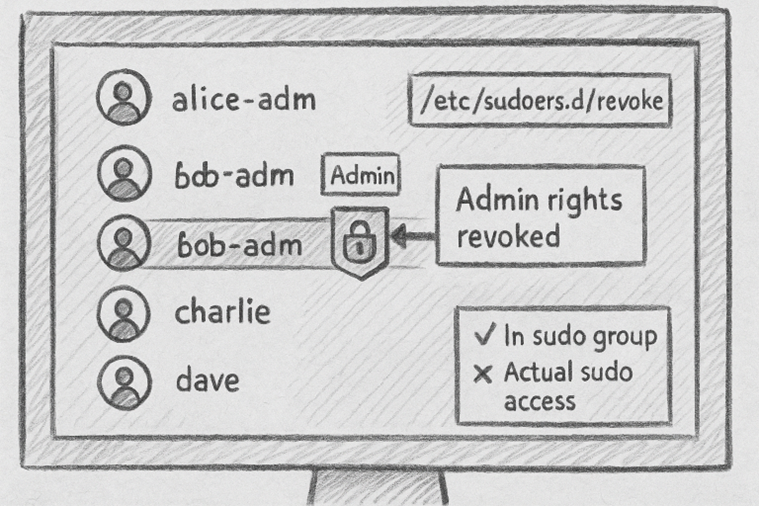}
  \caption{Trigger 2.1.1 (Base-Rate Neglect): multiple ``-adm'' accounts are visible during discovery while sudo privileges are revoked in configuration files, creating a choice between verification commands (e.g., \texttt{sudo -l}, \texttt{id}) and immediate escalation attempts. EKM sensors label verification as the rational path and immediate escalation as evidence of base-rate neglect. }
        \label{fig:BRN_illustration}
    \end{subfigure}%
    ~ 
    \begin{subfigure}[t]{0.48\textwidth}
        \centering
        \includegraphics[height=3cm]{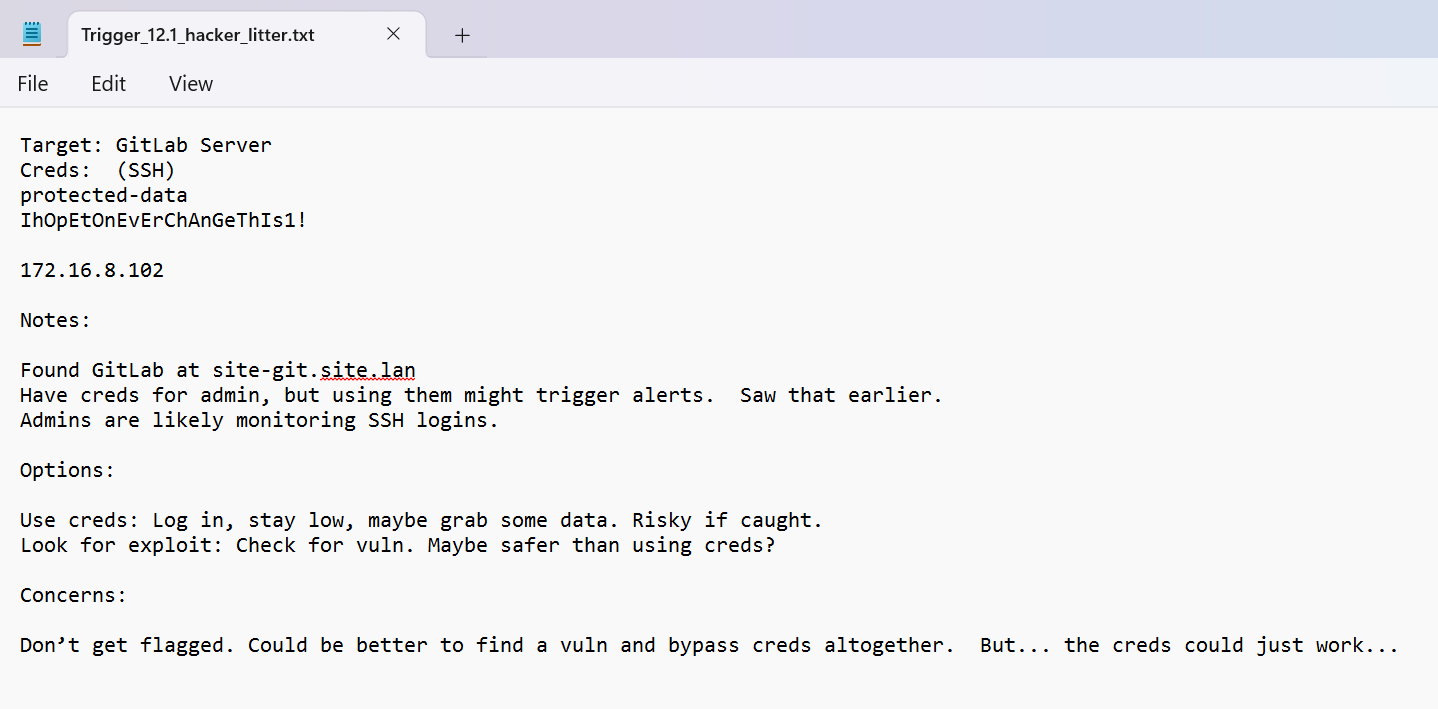}
        \caption{Trigger 12.1 (Loss Aversion): a \texttt{notes.txt} presents seemingly useful credentials alongside a slower, more reliable exploit route; sensors detect immediate credential use (SSH/SU attempts) versus methodical exploit preparation. Both triggers were instrumented for host/log-based sensing and produced measurable effects on time-on-task, diversion, and detectability (see Sec.~7).}
        \label{fig:LA_illustration}
    \end{subfigure}
    \caption{Examples of Triggers 2.1.1 (left) and 12.1 (right)}
\end{figure*}

This trigger was designed around Base Rate Neglect. It was located early on the attack path on \mt{it-ubuntu-1} in the \mt{site-it} segment. It contained multiple usernames ending with ``\mt{-adm}'' that were part of the \mt{sudo} group, creating the appearance of elevated privileges (as illustrated in Figure~\ref{fig:BRN_illustration}). However, both \mt{/etc/sudoers} and an entry in \mt{/etc/sudoers.d/} removed these privileges, meaning the accounts could not actually perform privileged actions. There was a larger number of these kinds of accounts than one might expect on a network of a similar size. The decision for the attacker was whether to trust the apparent group membership or verify actual permissions. This setup exploited Base Rate Neglect by encouraging attackers to act on the general expectation that accounts in the \mt{sudo} group have elevated rights, while ignoring the likelihood that these accounts might be exceptions to that rule.

\subsubsection{Trigger 12.1 (Loss Aversion)}
This trigger targeted Loss Aversion and was in the early part of the network on site-proxy in the site-it segment. A notes \mt{.txt} file on the system referenced potentially valuable credentials for protected-data user, offering a fast but uncertain path forward (see Figure~\ref{fig:LA_illustration}). An alternative mentioned in the notes file was to set up and execute a \mt{JNDI} style exploit on the \mt{Solr} service, which was slower but potentially more reliable. The attacker’s decision was whether to use the risky credentials immediately or pursue the more methodical exploit route. This design played on Loss Aversion by making the slower exploit path feel like a potential loss of time and opportunity, pushing attackers toward the immediate but riskier credential option to avoid that perceived loss.

\section{Expert Knowledge Model Sensors and Surveillance Sensors}

Although the CogVuln sensor provides a data-driven probabilistic estimate of attacker biases, it does so primarily through statistical inference over network and log data. To complement this automated reasoning, GAMBiT incorporates an additional layer of human-derived analytic models that capture expert understanding of how cognitive biases manifest in specific operational contexts. These \textit{Expert Knowledge Model (EKM) Sensors} translate qualitative behavioral expectations into formal, rule-based detection logic. Each EKM sensor encodes the reasoning of subject-matter experts (SMEs) who examined the triggers, observed attacker interactions, and defined measurable indicators of rational versus biased responses. Together, these sensors extend the CogVuln framework from purely inferential analysis to hybrid reasoning that fuses empirical data with domain expertise.

In parallel, GAMBiT employs a second class of analytic tools, \textit{Surveillance Sensors}, that monitor behavioral patterns across time and across triggers. Whereas EKM sensors are tightly coupled to specific cognitive triggers and their immediate decision contexts, surveillance sensors assess higher-order cognitive constructs such as risk tolerance, cognitive reflection, and adaptability. They analyze longitudinal features of attacker activity, including reaction times, tool-switching frequency, and sequence entropy, to infer broader tendencies and susceptibility to bias over the course of an engagement.

This section presents the design and implementation of both sensor classes. Section~6.1 introduces the Expert Knowledge Model Sensors, outlining their conceptual foundations, rule-construction methodology, and validation process. Section~6.2 then discusses the Surveillance Sensors, describing how cross-temporal behavioral analytics were used to capture stable cognitive traits and to complement the localized inferences of the EKM framework.

\subsection{Expert Knowledge Model Sensors}
\subsubsection{Overview and method}
The Expert Knowledge Model sensors are a type of post-trigger sensor, which leverages the fact that triggers induce a variety of behaviors that can be observed to sense bias-states in the attacker. Since this is a post-trigger sensor, this sensor is a key component to \GAM{}'s approach to design \Cyph{}s and \Aph{}s for continued bias sensing and to inform selection of defenses once triggers are encountered. To develop these sensors, cyber and cognitive psychology subject matter experts first defined each cognitive vulnerability trigger and its context: when it appeared in the scenario, the choices it created, and what actions tended to be more ``rational" or more ``biased" given the objectives. The team then turned those expert descriptions into plain-language expert-knowledge models (EKMs). Each EKM was translated into concrete code rules that identified where to look (hosts and log sources such as syslog and system.auth), what to exclude (training-platform artifacts), and which command or file-access patterns to count as indicative of the rational and biased decision paths. The rules were executed to generate per-participant counts by trigger and data source.

The team iterated on these rules across pilot and later sessions (Exp2 Control and Trigger). They tightened host scoping, added time windows around the cue, and cross-checked sources (for example, aligning syslog with system.auth) to cut false positives and improve trigger specificity. Version changes were tracked, and outputs were checked for platform artifacts and unexpected control-group activity. The result was a set of simple tables, counts by participant, trigger, and data source, ready for analysis across sessions and conditions.

\subsubsection{Sensor Examples}

\paragraph{B.2.1.1 (\mt{it-ubuntu-1} account/privilege cues)}

\begin{description}
    \item[Context.] While performing account discovery on \mt{it-ubuntu-1}, the attacker saw login names and group memberships that seemed to imply administrator (\mt{sudo}) privileges. At that moment, a more rational response might be to verify actual privileges first (e.g., \mt{sudo -l}, \mt{id}, \mt{groups}) before attempting escalation; a more biased response might be to assume privilege from surface cues and immediately try \mt{su}/\mt{ssh} escalation.
    \item[Data collection and interpretation.] The sensor read \mt{syslog} and \mt{system.auth}, filtered training-platform artifacts, and counted authentication/privilege events tied to trigger-specific accounts (e.g., ``\mt{-adm}'' conventions). Counts of \mt{su/ssh} activity linked to those accounts were treated as indicators of the biased path. Evidence of verification steps (e.g., \mt{sudo -l}, \mt{id}, \mt{groups}) was treated as an indicator of the rational path. 
\end{description}

\paragraph{L.12.1 (\mt{site-proxy} found-credentials cue)}

\begin{description}
    \item[Context.] While working on \mt{site-proxy}, the attacker found a notes file from a prior attacker describing potential credentials. If used they could create a tempting escalation shortcut but with higher detection/lockout risk. In that moment, a more rational response might be to treat the credentials as untrusted and either validate them carefully or pursue a deliberate exploitation path before use. A more biased response might be to try the found credentials immediately.
    
    \item[Data collection and interpretation.] The sensor queried \mt{syslog} for commands consistent with immediate credential use (e.g., \mt{ssh}, \mt{su} attempts referencing the found details) and checked \mt{system.auth} for corresponding session open/close events on the proxy. These patterns were used as indicators of the biased pursuit. In contrast, log evidence of methodical verification or exploitation (e.g., \mt{jndi -jar} or a comparable exploit string and related preparation) was used as an indicator of the rational path. 
\end{description}

\paragraph{Additional sensors}

Other sensors followed the same EKM-to-code approach: map each trigger’s decision points to concrete host/log signatures, exclude platform artifacts, apply the same iteration and quality-control steps, and produce per-participant counts that reflect the biased and rational paths consistent with each trigger’s framing.

\subsection{Surveillance Sensors}
In addition to the five aforementioned \CV{} sensors, we developed a surveillance sensor designed to detect more abstract constructs than the Cognitive bias-inspired \CV{}s. The surveillance sensor framework was grounded in cognitive psychology, behavioral economics, and decision science. It built on dual-process theories of reasoning, which distinguish between System 1 (fast, intuitive thinking) and System 2 (slow, deliberative thinking) \cite{Kahneman2012-qe,tversky1974judgment}. Under uncertainty or time pressure, attackers may rely on heuristics that can produce predictable cognitive biases, including loss aversion, base rate neglect, availability bias, confirmation bias, and the sunk cost fallacy. Three cognitive constructs formed the basis for inferring these biases: risk tolerance—the likelihood of taking high-consequence actions under uncertainty \cite{Weber2002-ru,Nicholson2005-ld,Grable2000-fr}; cognitive reflection—the ability to override initial responses and engage in effortful reasoning \cite{Frederick2005-zx,Toplak2011-hz,Pennycook2019-mw}; and flexibility—the capacity to shift strategies in dynamic environments \cite{Martin1995-fg,Diamond2013-sc,Stanovich2000-ng,Scott1962-ht}. Each construct was assessed through patterns in attacker behavior over time, using multiple data sources from the cyber range environment.

Risk Tolerance was designed to detect whether an attacker favored bold over cautious actions, consistent with research linking higher risk-taking to more intuitive, System 1–driven decision-making. Technically, the sensor parsed command line histories to flag early use of aggressive reconnaissance and exploitation tools (e.g., nmap with high-speed flags, hydra, msfconsole), correlated with Suricata alerts to detect noisy network activity soon after initial access, and examined host logs for rapid privilege escalation or access to sensitive files. Quick escalation attempts and high alert volumes early in the scenario were treated as indicators of higher risk tolerance, while delayed, low-noise approaches suggested lower tolerance for risk.

Cognitive Reflection was intended to identify the extent to which attackers engaged in structured, evidence-based reasoning, reflecting a shift from System 1 to more deliberative System 2 processing. The sensor evaluated command sequences for logical ordering, use of verification steps (e.g., chained commands with \mt{grep} or \mt{awk}), and adjustments following failed attempts. It also incorporated content from operation notes (e.g., ``Confirmed service availability'') and relevant self-report survey items. Repeated use of identical commands without variation, skipping verification, or acting on unconfirmed assumptions signaled lower reflection, while methodical adjustments and validation steps indicated higher reflection.

Cognitive Flexibility was designed, but not yet implemented, to measure adaptability in the face of failure or deception. The theoretical basis came from work on cognitive flexibility and problem-solving under dynamic conditions. Planned implementation included tracking when attackers abandoned ineffective approaches, switched tools, or explored alternative network paths after encountering misleading information. By correlating these behavioral shifts with trigger events in the network, the sensor would distinguish rigid, perseverative behavior from adaptive strategy changes.

The surveillance sensors were developed towards the end of the experiment and were not delivered as software.

\section{Experimental Results: Behavioral Impact}

\subsection{Mission Progress} 
\begin{figure}[htbp]
    \centering
    \includegraphics[width=1.0\linewidth]{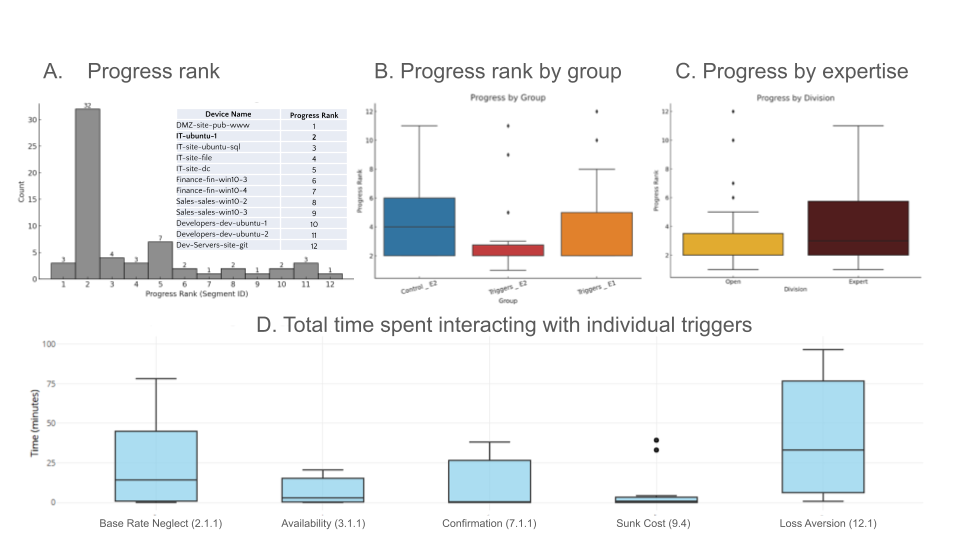}
    \caption[Progress and Trigger impact results]{Progress and Trigger impact results. A) Histrogram of progress made across participants, table inset indicates how progress rank was defined B) Progress rank by group C) Progress by expertise, comparing open vs expert division D) Estimated total time attackers spent actively interacting with individual triggers across a 16 hour experiment, shown for the 5 most encountered triggers.}
    \label{fig:Progress and Trigger Impact}
\end{figure}

Progress toward mission goals was evaluated by determining the deepest network segment each participant reached on a linear or shortest viable cyber-attack path. Each segment was assigned a rank from 1 (entry point) to 12 (deepest targets), providing an ordinal measure of how far participants advanced (see Figure~\ref{fig:Progress and Trigger Impact}, panel A). Most participants never made it past rank 2 (\mt{it-ubuntu-1}). Control participants in Experiment 2, who encountered no cognitive bias triggers, achieved greater median progress than both Trigger groups from Experiment 1 and 2, with a statistically significant difference (Kruskal-Wallis $p = 0.0495$;  see Figure~\ref{fig:Progress and Trigger Impact}, panel B. Expert division participants, defined by higher scores in a pre-experiment cybversecruity skills test, also progressed further than Open division participants (Mann-Whitney $p = 0.0424$; see Figure~\ref{fig:Progress and Trigger Impact}, panel C). These results suggest that the absence of bias-inducing triggers supported more focused, goal-oriented decision-making, while trigger exposure disrupted planning and execution. The skill-based performance gap underscores the influence of training and domain experience on success in complex cyber environments.

\subsection{Time wasted} \label{ch:results}
Some triggers were associated with significant time investment away from mission goals. Figure~\ref{fig:Progress and Trigger Impact}, panel D, shows the estimated time participants were actively interacting with each of the five most encountered triggers. These encounter times were derived from evidence of attackers interacting with triggers by pooling across both SOC-available and non-SOC available data. Although these trigger encounter times cannot be compared to the control group, this figure provides insight into time wasted by the attacker by choosing to interact with the trigger. Not all triggers were encountered by all participants, due to the open-world and naturalistic decision-making experimental design. The figure depicts the 5 most encountered triggers, showing that triggers 2.1.1 (base rate neglect) and 12.1 (Loss aversion) elicited the longest interactions times. 

\subsection{Diverting attackers off the attack path}
To evaluate whether triggers were diverting attackers off the attack path, the proportion of commands targeting a specific VM was calculated. To associate commands in the z-shell history logs to individual VMs, this metric assumes the first appearance of a specific VM name or corresponding IP in a command line as the “first” command interacting with that specific VM. Subsequent commands were associated with this specific VM until the appearance of another VM name \textbf{or} corresponding IP in a command line. By repeating this process for the entire shell history of each participant, the team associated every command with a VM and tallied the commands for each VM. Should the participants target more commands at the 13 VMs on the attack path, they were expected to achieve more hacking objectives or progress further, given fewer resources wasted on irrelevant VMs. Thus, to estimate attacker resources wasted, the proportion of commands on the attack path was calculated as follows:

\begin{equation}
    \frac{\text{Total number of commands targeted on the 13 critical VMs}}{\text{Total number of commands on all 38 VMs}}
\end{equation}

The equation above can also be formulated to examine the proportion of commands targeting a specific VM or group of VMs for analyzing trigger-specific behavioral impacts or particular time periods of the hacking exercise. Finally, specific commands were associated with \MA{} techniques, enabling a calculation of a proportion of \MA{} techniques targeting a specific VM. 
These three metrics are not independent from each other, which is why only a selection of key findings are presented here. 

\begin{figure}[htbp]
    \centering
    \includegraphics[width=1.0\linewidth]{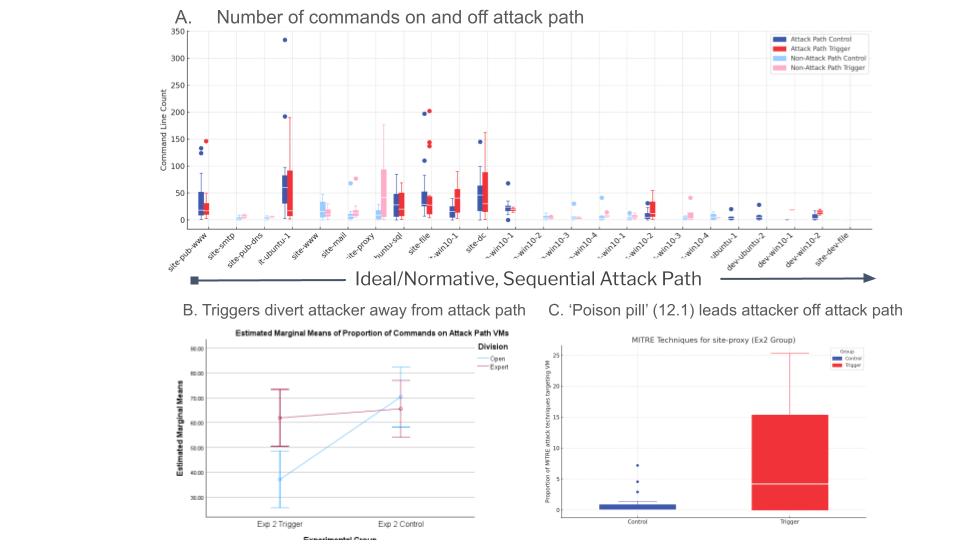}
    \caption[Triggers avert attackers off attack path]{Triggers avert attackers off attack path A) Number of commands for trigger and control groups per VM, VM's are ordered by their location on the attack path left to right). Solid colors are VM's on the attack path, faded colors are VM's not on the attack path. B) The proportion of commands on vs off the attack path for trigger and control groups, as well as open and expert divisions (blue and red colors) C) Trigger 12.1 leads attackers to site-proxy, this figure shows the proportion of TTP's targeting site-proxy for control and trigger groups}
    \label{fig:Attack path}
\end{figure}

Figure~\ref{fig:Attack path} shows the number of commands per VM for the trigger and control groups (panel A), with solid colors indicating VM's on the attack path. The proportion of commands on the attack path was shown to be lower for the control group (panel B), suggesting triggers collectively diverted attackers away from the attack path. A two-way ANOVA  reveals a significant main effect for Exp 2 Group (i.e., Trigger vs Control; F(1,35)=10.37, p = 0.003). The Exp 2 Control group has a greater portion of command lines targeting VMs on the attack path (Mean = 67.823; SD = 12.692) than the Exp 2 Trigger group (Mean = 49.501; SD = 24.256). The same two-way ANOVA also reveals a significant interaction effect between Exp 2 Group and Division (F(1,35)=6.37, p = 0.014).  Finally, one specific trigger (12.1; Loss Aversion) appears to have had a particularly large effect (panel C). This trigger diverts attackers to site-proxy, which shows a large increase in proportion of TTP's targeting this VM for the trigger group. A one-way ANOVA revealed a significant effect between Exp 2 Group on proportion of \MA{} techniques targeting VM “site-proxy” (F(1,37)=11.59, p < 0.01) and a significant main effect of Encountering Trigger L.12.1 (i.e., participants who encountered Trigger L.12.1 vs participants who did not encountered Trigger L.12.1 in trigger group + control group) on number of \MA{} techniques targeting VM “site-proxy (F(1, 37)= 5.67, p = 0.02). 
 
\subsection{Detectability of the Attacker}
\begin{figure}[htbp]
    \centering
    \includegraphics[width=1.0\linewidth]{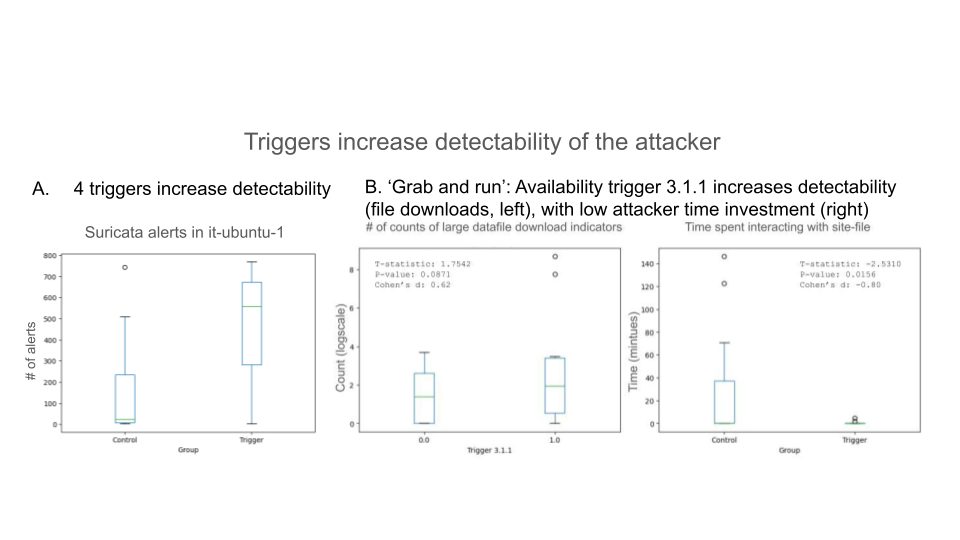}
    \caption[Triggers increase detectability of the attacker]{Triggers increase detectability of the attacker. A) Number of Suricata alerts on \mt{it-ubuntu-1} comparing control and trigger groups B) Activity on the site-file indicates Trigger 3.1.1 may induce `grab and run' activity, with evidence of increased counts of file download indicators (left) contrasting with low time investment interacting with the site-file (right) by attackers in the trigger group.}
    \label{fig:Detectability}
\end{figure}

Detectability of the attacker was measured by the number of Suricata alerts on \mt{it-ubuntu-1} (see  Figure~\ref{fig:Detectability}, panel A), which was higher for the trigger group compared to the control (T = 2.25, p = 0.0381). Three out of four triggers located on this VM appeared to contribute to increasing detectability according to multivariate analyses, taking the presence of triggers as independent variables (binary variable) and a z-scored value for Suricata alerts as dependent variable to make the coefficients similar to Cohen's d. Coefficients were positive for 2.1.1 (coeff = 0.35), 7.1.1 (coeff = 0. 53), 12.1 (coeff = 0.99), and negative for 9.4 (coeff = -0.85). 
Trigger 3.1.1 was located on a different machine, and could be analyzed separately (panel B). The figure shows that the number of file-download indicators was higher for people who encountered the triggers compared to the control, while attackers tended to spend fairly little time interacting with the VM on which the trigger was located. This suggests that trigger 3.1.1 may be a trigger that is best suited to increase detectability, but may not waste much time or attack resources. 

\subsection{Other trigger-specific behavioral impacts}
\begin{figure}[htbp]
    \centering
    \includegraphics[width=1.0\linewidth]{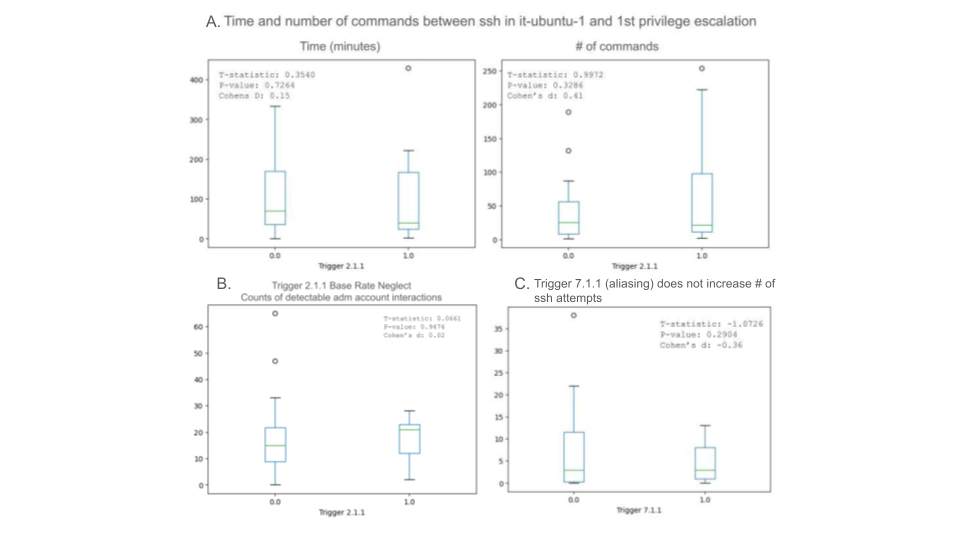}
    \caption[Other trigger-specific behavioral impacts]{Other trigger-specific behavioral impacts. A) shows the time (left) and number of commands (right) between when the attacker used \mt{ssh} to get into \mt{it-ubuntu-1}, and the first privilege escalation event. B) shows counts of detectable admin account interactions, and C) shows the number of times the attacker used SSH to get into \mt{it-ubuntu-1}, as Trigger 7.1.1 involved an aliased command that closes the session.}
    \label{fig:Trigger-Specific}
\end{figure}

Other trigger-specific behavioral impact analyses did not yield clear changes to the participants (see  Figure~\ref{fig:Trigger-Specific}). For Trigger 2.1.1, involving mislabeled admin accounts, no significant change in admin account interactions was detected, nor was the time or number of commands until first privilege escalation different between attackers who did and did not encounter the trigger. Moreover, for Trigger 7.1.1, aliasing a command that closes the attacker's session did not lead to an observable increase in the number of SSH attempts. These analyses underscore the challenge of behavioral impact analyses in a naturalistic decision-making experimental design, where effects of specific triggers that overlap in location and potential behavioral impacts, are not easily teased apart. 

\subsection{The GAMBiT approach: A rich dataset}
\begin{figure}[htbp]
    \centering
    \includegraphics[width=0.95\linewidth]{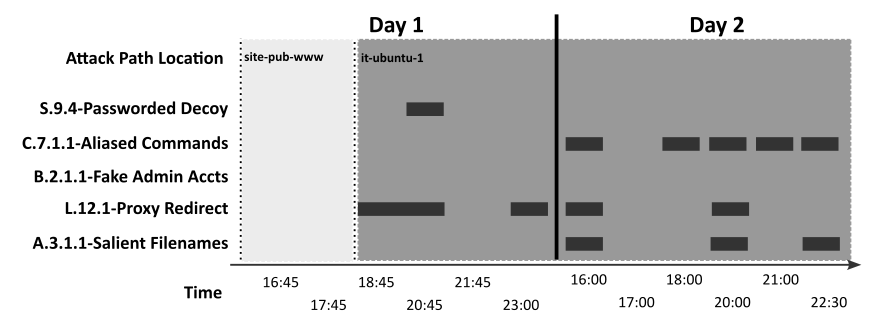}
    \caption[Illustration of triggers encountered]{Timeline of trigger encounters for an example participant across the two-day experiment. The horizontal axis shows time progression, while rows represent specific triggers encountered along the attack path, including S.9.4 (passworded decoy), C.7.1.1 (aliased commands), B.2.1.1 (fake admin accounts), L.12.1 (proxy redirect), and A.3.1.1 (salient filenames). Each horizontal block indicates a confirmed interaction period derived from command logs and network telemetry. The shaded regions separate Day~1 and Day~2, showing how trigger encounters varied by network location and task phase. The increased clustering of interactions on Day~2 suggests higher cognitive load and greater susceptibility to bias-driven behavior, illustrating the temporal dynamics captured by the CogVuln sensors.}

    \label{fig:example_trigger_encounters}
\end{figure}

The GAMBiT approach delivered a complex and rich dataset that is representative of real-world scenario's. One example of behaviors that could be observed in this approach is how attackers change their behaviors depending on context. Figure~\ref{fig:example_trigger_encounters} shows trigger interactions for one participant over the course of the 16 hour experiment. This particular participant never made it past rank 2 (\mt{it-ubuntu-1}), and appears to have increasing trigger interactions towards the end of the experiment. There may be different explanations for such behavior, such as participants not encountering triggers until later. However, another hypothesis is that time-pressure, frustration, and fatigue may build up during the experiment, inducing an increased vulnerability in the participant to fixed on their biased behaviors that has been observed in major accidents, such as Three Miles Island. This hypothesis is a key component of GAMBiT's original approach that aimed to overwhelm the frontal cortex (e.g., rational thinking) with emotion, stress, and other factors, and drive the attacker towards more biased or ineffective behaviors. However, more research is needed to confirm the effectiveness of this approach.

\section{Conclusion and Future Work}

By embedding cognitive science into cybersecurity, we demonstrate how attacker reasoning can be manipulated to the defender’s advantage. Our findings open a pathway for next-generation cyber defense systems that exploit adversarial cognition, predict attacker behavior, and adapt defenses dynamically. Future work will extend toward personalized cognitive modeling and autonomous orchestration of defense policies.

\backmatter

\bmhead{Acknowledgements}

This work was supported by the Intelligence Advanced Research Projects Agency (IARPA) under the Resilient Security and Control for Intelligent Networked Defense (ReSCIND) program. The authors gratefully acknowledge the contributions of the GAMBiT/ReSCIND performer team for their efforts in scenario design, experiment execution, data curation, and analysis. We thank the SimSpace Cyber Force Platform operations staff for range support and technical assistance throughout data collection. 

\section*{Declarations}

\subsection*{Funding}
This research is based upon work supported in part by the Office of the Director of National Intelligence (ODNI), Intelligence Advanced Research Projects Activity (IARPA), via N66001-24-C-4504. The views and conclusions contained herein are those of the authors and should not be interpreted as necessarily representing the official policies, either expressed or implied, of ODNI, IARPA, or the U.S. Government. The U.S. Government is authorized to reproduce and distribute reprints for governmental purposes notwithstanding any copyright annotation therein. Additional institutional support for this work came from the Bulls Run Group, Raytheon Technologies, the University of Southern California Institute for Creative Technologies, Northeastern University, Virginia Tech, and New York University.

\subsection*{Conflict of interest / Competing interests}
The authors declare that they have no known competing financial interests or personal relationships that could have influenced the work reported in this paper.

\subsection*{Ethics approval and consent to participate}
All human-subject data were collected under Institutional Review Board (IRB) approval from Virginia Tech (IRB \#23-565). All participants provided informed consent prior to participation, acknowledging that their de-identified data would be shared for scientific and research purposes under controlled-access repositories.

\subsection*{Consent for publication}
All participants consented to publication of de-identified, aggregate data derived from their experimental sessions and self-reports.

\subsection*{Data availability}
The datasets generated and analyzed during the current study are publicly available on IEEE DataPort:
  \begin{itemize}
    \item Experiment 1 — \url{https://dx.doi.org/10.21227/dwkg-n940}
    \item Experiment 2 — \url{https://dx.doi.org/10.21227/39z8-w554}
    \item Experiment 3 — \url{https://dx.doi.org/10.21227/xdw9-3677}
  \end{itemize}
  All datasets include documentation, metadata, and curated derivative products as described in Section~2.

\subsection*{Materials availability}
All experimental materials, including trigger code templates, network topology documentation, and sensor schemas, are available on a reasonable request from the corresponding author, subject to institutional data sharing policies and export control.

\subsection*{Code availability}
Custom data-processing scripts used for cleaning, alignment, and feature extraction (Python-based parsers for Suricata and host logs, and CogVuln sensor prototype modules) are available at the NYU Secure Systems Lab GitHub repository (access upon request).

\subsection*{Author contribution}
All authors contributed to the conceptualization, experimental design, data collection, curation, and analysis of the GAMBiT datasets. Brandon Beltz, Jim Doty, and Rachelle Thomas led experimental coordination; Yvonne Fonken, Bret Israelsen, and Peggy Wu supervised data engineering and cognitive-bias instrumentation; Stacy Marsella and Nikolos Gurney contributed to behavioral modeling and cognitive analysis; Nathan Lau and Stoney Trent oversaw methodology validation; Ya-Ting Yang and Quanyan Zhu provided theoretical modeling, data synthesis, and manuscript preparation. All authors reviewed and approved the final manuscript.

\bibliography{cyp2025refs}

\end{document}